\documentclass[11pt]{article}
\usepackage{latexsym}
\usepackage{amsmath}
\usepackage{multicol}
\usepackage{moreverb}
\usepackage{alltt}
\usepackage[algosection, boxed]{algorithm2e}
\usepackage[a4paper, lmargin=1.5in, rmargin=1.2in, tmargin=1.5in, bmargin=1.0in]{geometry}

\newcommand{\rar}{\rightarrow}

\newtheorem{theorem}{Theorem}[section]

\newcommand{\qed}{\nobreak \ifvmode \relax \else
      \ifdim\lastskip<1.5em \hskip-\lastskip
      \hskip1.5em plus0em minus0.5em \fi \nobreak
      \vrule height0.75em width0.5em depth0.25em\fi}

\begin{document}


\title{Techniques for Distributed Reachability Analysis with Partial Order and Symmetry based Reductions}
\author{{\bf Janardan Misra} \\       
        HTS (Honeywell Technology Solutions) Research Lab\\
151/1 Doraisanipalya, Bannerghatta Road,\\ Bangalore 560 076, India\\
{\sf Email: janardan.misra@honeywell.com}\\
\and
        {\bf Suman Roy}~\thanks{Work done when authod was associated with HTS.}\\
SETLABS, Infosys Technologies Ltd.\\
\#44 Electronic city, Hosur Road, \\ Bangalore 560 100, India.\\
{\sf Email: suman\_roy@infosys.com}
}
\maketitle

\begin{abstract}
In this work we propose techniques for efficient reachability analysis of the state space (e.g., detection of bad states) using a combination of partial order and symmetry based
reductions in a distributed setting. The proposed techniques are focused towards explicit state space enumeration based  model-checkers like SPIN. We consider variants for both depth-first as well as breadth-first based generation of the reduced state graphs on-the-fly.
\end{abstract}
\tableofcontents

\section{Introduction}

State space explosion is a fundamental bottleneck in verifying
large scale industrial systems (software and hardware) using model checking
methods~\cite{CGP00,BBFLPPS01}. Primarily the problem is the result of 
thrashing owing to excessive page faults as the state space becomes too large to be contained in the main memory.
There are several approaches to (partially) overcome the
problem. Partial order and Symmetry based methods for state space 
reduction are the most important among them. Distributed verification 
framework is yet another important measure to overcome the memory limitations.

Partial order based methods exploit the independence of actions to
reduce the size of the state space. The basic idea is that given a set of interleaving
sequences of actions one can define sequences that are equivalent
upto reordering of independent actions. For all those
specifications, which do not distinguish between the equivalent
sequences, one can consider a representative subset of sequences from each
equivalent class. This generates a reduced state space including
only a subset of sequences. Most methods work by exploring a
subset of the actions enabled from a state. The subset is selected
according to some constraints that guarantee that enough
representatives, at least one from each equivalent class, will be
generated.

On the other hand symmetry based methods exploit the architectural
symmetry present in the system description. Finite state
concurrent systems frequently contain replicated components. For
example, a network protocol might involve a large number of
identical processes communicating in some fashion. Hardware
devices also contain memories with replicated components. Thus knowledge of the 
presence of identical components can be 
 used to generate reduced models of these systems because
presence of symmetry in the system induces a transition structure
preserving equivalence relation on the state space. Thus while
performing model checking one can discard a state if another
equivalent state has already been considered.

Distributed methods basically aim toward distributing the memory
requirement for storing the state space on many processing
nodes on a network/cluster without incurring high communication
overhead. Often this distribution of model
checking process actually helps in overcoming the memory
limitations for large state spaces rather than scaling the
processing time.

In this invention, we propose a design framework for efficient model
checking of the reachability properties (e.g., detection of bad
states) using a combination of partial order and symmetry based
reductions in a distributed setting. We consider variants for both depth
first traversal and breadth first traversal.

\section{Background}

\subsection*{Preliminaries}

A \emph{Labeled Transition System} (LTS) is a tuple $(S, R, Act, s_0, L)$, 
where $S$ is a finite set of states and $s_0 \in S$ is an initial state. $Act$ is a (possibly infinite) set of action-labels. $R \subseteq S \times Act \times S$ is a transition relation such that $(s, \alpha, s') \in R, \alpha \in Act$
is also written as $s \stackrel{\alpha}{\rightarrow} s'$.   $L:S\rightarrow 2^{AP}$ is the state labeling function such 
that $\forall s \in S$, $L(s)$ defines the
subset of propositions true in $s$. For example, if $v$ is a
boolean variable live in state $s$, then $v \in L(s)$ indicates
that when system reaches state $s$, $v = True$ and $v \not\in
L(s)$ denotes that $v=False$. $AP$ is the set of all atomic
propositions. Let $(\forall s \in S)\ enabled(s)=\{\alpha
\in Act \mid \exists s' \in S\  s.t.\ s \stackrel{\alpha}{\rightarrow} s'\}$, 
be the set of all the transitions enabled in a state $s$. In case of deterministic processes, we write, $s' = \alpha(s)$ for $s \stackrel{\alpha}{\rightarrow} s'$ and for non deterministic processes, we write, $s' \in \alpha(s)$ 
for $s \stackrel{\alpha}{\rightarrow} s'$. In the following discussion, we will only consider deterministic processes, though the discussions can be smoothly extended to the case of non deterministic processes. 

In practice, a typical system model may contain several concurrent 
processes such that a {\it global state} refers to the valuations for all 
variables and {\it local state} of a process is restricted to the valuations 
only for those variables accessible to the process. 

\subsection{Centralized DFS Algorithm}
\label{basic-dfs}

We now recall the basic Depth First Search (DFS) algorithm used to
explicitly construct the state graph of a given transition system: 

\begin{verbatim}
// Centralized DFS Based State Space Generation

gen_state_graph() {
  V := {} /* set of generated states */
  Stack U := {} /* stack of states */
  dfs(Start_state)
}

dfs(s) {
   if (s is not in V)
   V := V + {s}
   push_stack(U, s)
   for each (sequential process P)   
     nxt := all transitions of P enabled in s
     for each (t in nxt)
        st := t(s)
        dfs(st)         
   pop_stack(U) 
}
\end{verbatim}

We assume a system model containing several concurrent processes. 
The stack (U) contains the states generated while traversing 
from the initial state to the current state. The set of visited states V
is implemented by a hash table with collision lists, which is
generally the most memory consuming data structure, its size being
proportional to the number of states in the state graph. Also the
stack data structure can consume a considerable amount of memory,
because its size is proportional to the depth of the state graph,
which, in some cases, can be comparable with the number of states. 
\subsection{Centralized BFS Algorithm}

Basic implementation of Breadth First Search (BFS) based state space generation 
is not much different from DFS. Instead of a search stack, it uses a queue, which allows
insertion and deletion operations in first-in-first-out basis.
Successor states are inserted  to the tail of the queue
(add\_q(Q, s)) and states for expansion are removed from the
front of the queue (pop\_q(Q)). 
\begin{verbatim}
// Centralized BFS Based State Space Generation

gen_state_graph() {
  V := {} 
  Queue Q = {} /* Queue of states to be expanded. */
  start(Start_state);
}

start(s) {
  add_q(Q, s)
  V = V + {s}
  bfs(Q)
}

bfs(Q) {
  s = pop_q(Q)
  for each (sequential process P)
     nxt := all transitions of P enabled in s
     for each (t in nxt)
       succ = t(s) /* t(s) is the state reached on  
                      taking transition t from state s */
       if (succ is not in V)
          add_q(Q, succ) /* Insert succ at the tail of Q */
          V = V + {succ} /* Add succ to State graph V */        
   if (Q != {})
   bfs(Q)
}
\end{verbatim}

Unlike DFS, BFS cannot be extended smoothly to detect cycles.
Again contents of queue $Q$ are not sufficient for traversing an
error path and more information such as a link to parent node is
required to trace any such error path. Nonetheless, if an error
state is encountered, BFS yields shortest path from the start
state. 

\subsection{Basic Distributed DFS Algorithm}
\label{dis_dfs}

For the distributed computation we assume a network of
collaborating processors with no global memory under MIMD (multiple input multiple data) 
architecture. Communication among these processors is realized by means of message passing
only. The basic idea behind the parallelization (distribution) is to
divide the generation of the state graph into
independent subtasks that can be performed in an arbitrary order
in parallel. This is achieved by dividing state transition system
into partitions, one partition for each processor. In practice, it results into
splitting the search stack into parts determined by fully expanded 
states such that each processor operates only on its part - 
the local search stack ($V[i]$). Thus every node in the network 
owns one of the state subsets, and is responsible for
holding the states it owns and for computing their successors.
When a node computes a new state, it first checks if the state
belongs to its own state subset or to the subset of another node.
If the state is local, the node goes ahead as usual, otherwise a
message containing the state is sent to the owner of the state.
Received messages are held in a queue and processed in sequence.
distributed termination algorithms, e.g., ring algorithms~\cite{} can be used 
when all queues are empty and all nodes are idle to terminate the verification.

The following pseudo-code illustrates the DFS based algorithm used in the
distributed version as described in~\cite{LS99}:
\begin{verbatim}
// Distributed DFS Based State Space Generation

Search(i, Start_state) {
   V[i] := {} /* Set of already visited states on processor i */
   U[i] := {} /* Local stack at processor i */   
   j := partition(Start_state)    
   if (i == j)
      push_stack(U[i], Start_state)
   visit(i)
 }

visit(i) {
  receive(); 
    /* receive() accepts states generated at other processors to 
       be owned by this processor. It adds these states to U[i]. */ 
    
  while(True)
    while (U[i] != {})
       s: = top_stack(U[i])
       dfs(i, s)       
}

/* Each processor i on the network computes its own local search space. */

dfs(i, s) {
  if (s is not in V) /* For this we need to check for all the state-lists
                        across (all) processors in the network.*/
    V[i] := V[i] + {s}
    push_stack(U[i], s)
    for each (sequential process P)
      nxt := enabled transitions of P in s
      for each (t in nxt)
        st := t(s)
        j  := partition(st)
        if (i == j)
           dfs(i,st)
        else
           send(st, j)                   
    pop_stack(U[i])  
}   
\end{verbatim}

It is assumed that all  the processors on the  network are given
unique  integer   identifiers  at the  beginning. And  function
partition(s) returns the identifier of the processor to own the
state s. visit() continually looks for received messages containing states
and their history and adds these states to local queue.  

\subsection{Partial Order Reduction} \label{sec:po}

As discussed before, the state space explosion is one of the
fundamental bottlenecks for applying model checking algorithms on
large complex systems. In case of asynchronous systems with large
number of concurrently active processes, the state space explosion
problem arises out of the interleaving semantics implying all
possible interleaved ordering of transitions of these processes. 
In fact, for a set of just $n$ transitions, which can
be executed concurrently, there are $n!$ different ordering and
$2^n$ different states. Now if the specification does not
distinguish between these different orderings then only one
ordering with just $n+1$ states is sufficient to consider. This is
what is essentially exploited in the method of partial order
reduction (POR)~\cite{HP94}. To achieve this, POR methods 
exploit the commutativity of concurrently executed
transitions, which result in the same state when executed in
different order. Thus it is primarily suited for asynchronous
systems/protocols since in synchronous systems concurrent actions
are executed simultaneously rather than being interleaved.

The POR can be achieved by modifying the basic DFS
procedure as described in Section~\ref{basic-dfs}. The search starts with the
initial state $s_0$ and proceeds recursively. For each generated state $s$
it selects only a subset $ample(s) \subseteq enabled(s)$ of enabled transitions, yielding reduced state graph.

\begin{verbatim}
// Partial Order Reduction with basic DFS 

Search() {
  V := {} /* State Graph - set of visited states */
  Stack U := {} /* DFS stack of states */
  dfs(Start_state)
}

dfs_po(s) {
 if (s is not in V)
   V := V + {s}
   push_stack(U, s)
   for each (sequential process P)
     nxt := all transitions of P in ample(s)
     for each (t in nxt)
        st := t(s)
        dfs(st)
   pop_stack(U)
}
\end{verbatim}

The main step of the algorithm is to formulate an effective
and efficient way to determine $ample(s)$ for any given state $s$ such that
the verification result must be the same for the reduced and original graph.
This is achieved by defining function $ample()$ in accordance with the 
correctness preserving conditions that are commonly known as conditions \textbf{C0-C3} (\cite{CGP00}.)
We will discuss only conditions \textbf{C0-C2}, which are sufficient to verify the safety properties. 
Further details on these conditions can be found in~\cite{CGP00}. 

\begin{itemize}
    \item \textbf{C0}: $(\forall s \in S)ample(s)= \emptyset \Leftrightarrow
enabled(s) = \emptyset$
\end{itemize}

This condition guarantees that if a state has a successor then
reduced state graph will also have a successor for this state. 
As a consequence this constraint ensures that presence of a deadlock in the reduced graph 
also implies a deadlock in the original state graph.

\begin{itemize}
    \item \textbf{C1}: For each state $s \in S$, along every path in the full
state graph that starts at $s$, if there exists some action $\beta
\in Act$, which is dependent on some action appearing in
$ample(s)$, then $\beta$ cannot be executed without a
transition in $ample(s)$ occurring first.
\end{itemize}

Note that this condition is not immediately checkable by examining
only the current state since it refers to future states in the
full state graph, which might not even be present in the reduced
graph. In worst case it might demand constructing the full state
graph since checking it turns out to be as hard as solving the
Reachability problem for the full state graph. Therefore in
practice we avoid checking \textbf{C1} for all arbitrary subsets
of enabled transitions and instead use heuristics such as the one
presented in~\cite{CGP00}, though we might not always
achieve the optimal reduction.

Condition \textbf{C2} is based upon the observation that if there exist 
a cycle in the reduced state graph and there exist some action which 
was enabled in a state along the cycle in
the full state graph but was not considered in any of the ample
sets for the states in the cycle, then it will get permanently
ignored \emph{(ignoring problem)} in the reduced state graph. To
avoid such ignoring of transitions we consider the following
condition:

\begin{itemize}
    \item \textbf{C2 (Cycle proviso)}: For every cycle in the reduced state
graph, say $C= s_0 \stackrel{\alpha_0}{\rightarrow} s_1
\stackrel{\alpha_1}{\rightarrow} s_2 \ldots s_n
\stackrel{\alpha_n}{\rightarrow} s_0$, we have $\bigcup_{s_i \in
C}ample(s_i)=\bigcup_{s_i \in C}enabled(s_i)$.
\end{itemize}

In practice cycle proviso is efficiently implemented with
respect to the specific traversal algorithm used for
generating the state space. In the usual DFS based scenario
following stronger proviso is instead checked:

\begin{itemize}
    \item \textbf{C2* (Cycle proviso for DFS)}: $(\forall s \in S)\ ample(s)
\subset enabled(s) \Rightarrow \ \not\exists \alpha \in ample(s)\
s.t. \alpha(s) \in U$, where $U$ is the list of states in the DFS
stack.
\end{itemize}

The above three conditions are sufficient to verify any
\emph{equivalence robust} property (see~\cite{KLMPY02} for
details), which include both safety as well as fairness properties. 
Determining optimal ample sets for all the
states in an arbitrary state graph is a hard problem (with respect
to space requirements) particularly when we aim to calculate these
ample sets per state, on-the-fly, while generating the graph.
Therefore, in practice, heuristics are used
instead to check the conditions. Below we present one such
heuristic discussed in~\cite{CGP00} for \textbf{C2/C2*}. 

\begin{verbatim}
// Heuristic for ample()

ample(s) {
   for each (sequential process Pi such that Ti(s) != {})
       if( check_C1(s, Pi) AND check_C2*(s, Ti(s)) 
           AND check_C3(s, Ti(s)))
         return Ti(s)  
   return enabled(s)
}

check_C2*(s, X) {                  
  for all (t in X)                                                  
      if (t(s) in V)                                        
          return False                                                                                            
  return True                    
}                                 
\end{verbatim}

However, if we consider a concurrent system as a parallel composition of
sequential processes, an obvious candidate for $ample(s)$ is the
set $Ti(s)$ of transitions enabled in $s$ for some process $Pi$.
Because the transitions in $Ti(s)$ are interdependent, an ample
set for $s$ must include either all of the transitions or none of
them.  To construct an ample set for the current state $s$, we
start with some process $Pi$ such that $Ti(s) \neq \emptyset$. We
want to check whether $ample(s)=Ti(s)$ satisfies all conditions
\textbf{C1, C2/C2*}, and \textbf{C3}. If either of these remain
unsatisfied for all the processes active in state $s$, we take $ample(s)=enable(s)$.

\subsection{Distributed DFS Algorithm with Partial Order Reduction}

A distributed version of the DFS based algorithm for PO
reduction was presented in~\cite{BCMS05,PG02}. We will discuss their
approach in this section, which will be augmented with symmetry
based reductions later. The key part of such an algorithm is the distributed
checking of the ample conditions.

While conditions \textbf{C0} and \textbf{C1} can be checked
locally, checking condition \textbf{C2/C2*} requires special 
consideration in a distributed environments.
Therefore aim is to define a counter part of the condition \textbf{C2*}
for the DFS based generation of the state transition system which
is distributed among several processors. 
Motivated by the observation that
during the DFS search only a part of the search stack is needed in
order to ensure the condition \textbf{C2*}, authors propose a variant of 
\textbf{C2*}. In particular, since the relevant part of the search stack 
lies between the top of the stack and the topmost state that has been 
fully expanded. This
is because after a state has been fully expanded (i.e.,
$ample(s)=enabled(s)$) all the cycles reaching this state through
the search stack contain this state as a fully expanded one. Based
on this simple observation authors suggest to split the reduction
(generation) process into independent subtasks such that each time
a state is fully expanded, a new search with an empty search stack
is started. This is particularly suitable for distribution since
it does not need to care about transferring search stacks among
the processors. Several subtasks can be performed in parallel on
different processors. To deal with ``global cycles" (stretching
over more than one processor), fully expand a state whenever
crossing to a different processor.

Thus, each node maintains a stack states from which the generation
of the reduced state transition system is to be started. A
\emph{manager} initiates the entire computation by starting the
first DFS procedure from the initial state. Whenever a new state
$s$ is visited a set $ample(s) \subseteq enabled(s)$ of
transitions is computed. Always a set that fulfills the ample
conditions is selected; in particular it does not include a
transition leading to a search stack nor to another node. If such
a set can not be found, the current state is fully expanded.

There are two possible scenarios. In case the state $s$ is
fully expanded, every successor $s'$ of the state $s$ is inserted
into the stack. If $owner(s')$ differs from $owner(s)$ a message
is sent to the owner of $s'$ to do so. The DFS then backtracks
from the state $s$. Otherwise, the DFS continues generating the
state transition system following transitions from $ample(s)$
only. After the DFS ends, all incoming messages are processed.
Then a state from the stack is picked and a new DFS is initiated
from it. This step is repeated until the stack is empty. Once the
stack is empty and there are no incoming messages, the node starts
to \emph{idle}. If all nodes are idle and there are no pending
messages the algorithm terminates.

The modified algorithm for computing ample set in distributed
setting is presented as following: Note that only change is required in 
\emph{check\_C2*()}.

\begin{verbatim}
// Modified Heuristic for C2* 

ample(s) {
   for each (sequential process Pi such that Ti(s) != {})
       if ( check_C1(s, Pi) AND check_C2*(s, Ti(s)) 
            AND check_C3(s, Ti(s)))
          return Ti(s)       
   return enabled(s)
}
   .
   .
   .
check_C2*(s, X){
  for all (t in X)
     if ( (t(s) is in U[i]) OR   
          (partition(t(s)) < i) ) 
        return False        
  return True
}
\end{verbatim}

For improving the efficiency of cycle detection a state is sent to the owner node only if the id of the owner is greater than the id of the current node. Notice hat now $check_C2*(s, X)$ would detect a cycle if there exist some successor of state $s$, which is already present in the local search stack $U[i]$ OR the id for the owner of successor state is strictly less than the id for the current node i - this is because - \textbf{\textit{such a state would imply that there must exist some other state on the path from the start state to the current state which would have been visited by some other node earlier, hence closing the cycle. }}

Based upon the distributed version of determining \textbf{C2*}, we
can describe the complete distributed DFS algorithm with POR as follows:


\begin{verbatim}
// Distributed DFS algorithm with Partial Order reduction

Search(i, Start_state) { .. }

visit(i) {
  while(True)
    while (U[i] != {})
       s: = extract(U[i])
       ddfs_po(i, s)  
}

ddfs_po(i,s) {
  /* we need to check if state s is already generated */
  if (s is not in V)
     V[i] := V[i] + s
     push_stack(U[i], s)
     for each (sequential process P)
        nxt := all enabled transitions of P in ample(s)
        for each (t in nxt)       
          st := t(s)
          j  := partition(st)
          if (i == j)
             ddfs_po(i, st)
          else
             send(st, j)         
     pop_stack(U[i])
}
\end{verbatim}

\subsection{Symmetry in Model Checking}

Yet another way to combat the state space explosion problem 
is to exploit symmetries in a system description~\cite{CJEF96,DBH03,ES96,I05}. 
To illustrate, consider a mutual exclusion protocol based on
semaphores. The (im)possibility for processes to enter their
critical sections will be similar regardless of their identities,
since process identities (pids) play no role in the semaphore
mechanism. More formally, the system state remains behaviorally
equivalent under permutations of pids. During state-space
exploration, when a state is visited that is the same, up to a
permutation of pids, as some state that has already been visited,
the search can be pruned. The notion of behavioral equivalence
used (bisimilarity, trace equivalence, sensitivity to deadlock,
fair- ness, etc.) and the class of permutations allowed (full,
rotational, mirror, etc.) may vary, leading to a spectrum of
symmetry techniques.

The two main questions in practical applications of symmetry based
techniques are how to find symmetries in a system description, and
how to detect, during state- space exploration, that two states
are equivalent. To start with the first issue: as in any other
state-space reduction method based on behavioral equivalences, the
problem of deciding equivalence of states requires, in general,
the construction of the full state space. Doing this would
obviously invalidate the approach, as it is precisely what we are
trying to avoid. Therefore, most approaches proceed by listing
sufficient conditions that can be statically checked on the system
description. The second problem, of detecting equivalence of
states, involves the search for a canonical state by permuting the
values of certain, symmetric, data structures. In~\cite{CJEF96} it
was shown that the problem is at least as hard as testing for
graph isomorphism, for which currently no polynomial algorithms
are known. Furthermore, this operation must be performed for every
state encountered during the exploration. Solutions
proposed in the literature either deal with incomplete equivalence
classes for which the canonicalization problem has polynomial
solution~\cite{CJEF96} or use heuristic strategies~\cite{DBH03,I05}.

A common heuristic approach is to use a representative function, 
$rep : S \rar S$ that, given a state $s$,
returns an representative state state from the equivalent class containing $s$, 
induced by symmetry transformation on the system.  For an equivalence
class $C$, the states in $\{rep(s) \,|\, s \in C\}$ are called the
{\it representatives} of $C$. Clearly, if $rep(s) = rep(t)$ then
states $s$ and $t$ are equivalent. The reverse does not hold in
general, but the smaller the set of all representatives of a
class, the more often it will hold. 
The definition of the representatives is usually based on some
partial ordering on states, e.g. by taking as representatives
those states that are minimal in a class, relative to the
ordering. If the ordering is total, then the representatives are
unique for their class.

%

\section{Extended Distributed DFS Algorithms}

\subsection{A Distributed DFS Algorithm with Symmetry Reduction}

Assuming $rep(s)$ is available, we can modify the distributed DFS based state
space generation algorithm presented in the Section~\ref{dis_dfs} as follows: 
\begin{verbatim}
// Distributed DFS algorithm with Symmetry reduction 

Search(i, Start_state) { ... }

visit(i) { ... }

dfs(i,s) {
  if (rep(s) is not in V)
    V[i] := V[i] + {rep(s)} /* Note that we store only rep(s), not s. */
    push_stack(U[i], s)
     .
     .
     . 
}
\end{verbatim}

\begin{theorem}[Correctness] \label{th1}
The reduced state graph generated by the above algorithm is equivalent to the
full state graph under reachability properties. 
\end{theorem}
{\bf Proof Sketch.}
Follows from the correctness of basic distributed DFS algorithm and function $rep()$. This is based upon the observation 
that steps in the basic distributed DFS algorithm do not distinguish between state $s$ and $rep(s)$. Therefore algorithm would generate symmetry reduced graph consisting of states returned by $rep()$. $\qed$

\subsection{A Distributed DFS Algorithm with both Partial Order and Symmetry Reductions}

We can extend the basic DFS algorithm presented in the Section~\ref{sec:po} 
to handle both partial order and symmetry based reductions in a distributed setting. 

\begin{verbatim}

// Distributed DFS algorithm with both 
   Partial Order and Symmetry Reductions //

Search(i, Start_state) { ... }

visit(i) { ... }

dfs_po_symm(i, s) {
  /* we need to check if rep(s) is already in 
     the stack across (all) processors.*/
  
  if (rep(s) is not in V)
     V[i] := V[i] + rep(s) /* Note that we store only rep(s), not s. */
     push_stack(U[i], rep(s)) 
     for each (sequential process P)     
        nxt := all transitions of P in ample(s)
        for (each t in nxt)       
          st := t(s)
          j := partition(st)
          if (i == j)
            dfs_po_symm(i,st)
          else
            send(j, st) /* Send the state st to its owner j */                   
     pop_stack(U[i])  
}

check_C2*(s, X){
  for all (t in X)
     if ( (rep(t(s)) is in U[i]) OR
          (partition(rep(t(s))) < i) )     
        return False
  return True
}
\end{verbatim}
Note that $Check\_C2*()$ returns True only if rep(t(s)) is already present in local 
search stack  OR the id of the owner of the representative of the successor state is less than the id of the the current node.

\begin{theorem}[Correctness]\label{th2}
The reduced state graph generated by the above algorithm is equivalent to the
full state graph under LTL$_{-X}$ properties including reachability properties. 
\end{theorem}
{\bf Proof Sketch.} It was proved in~\cite{EJP97} that LTS generated using $(ample\circ rep)()$ is equivalent to full LTS under LTL$_{-X}$ properties including safety properties. Together with the correctness claim in Theorem~\ref{th1}, correctness of the algorithms follows. $\qed$

\section{Extended Distributed BFS Algorithms}

\subsection{A BFS Algorithm with Partial Order Reduction}

We can modify the basic BFS algorithm for POR by limiting 
the transitions per state to the ample sets. These ample
sets can be calculated using the heuristics presented in
Section~\ref{sec:po}.

\begin{verbatim}
//Distributed BFS with Partial Order Reduction 

Search(i, Start_state) {
   State Space V[i] := {} /* List of visited states for processor i */
   Queue Q[i] := {}       /* local queue at processor i */
   j := partition(Start_state) 

  if (i == j)
     add_Q(Q[i], Start_state)
     visit(i)
}

visit(s) { ... }

check_C2*(s, X){
  for all (t in X)
     if ( [t(s) is in V[i]) AND (t(s) is not in Q[i])] OR
          [partition(t(s)) < i] )
        return False
  return True
}

bfs_po(Q, i) {
  s = pop_q(Q[i]);

  for each (sequential process P)
    nxt := all transitions of P in ample(s)
    for each (t in nxt)
      succ = t(s)
      if (succ is not in V)
         j := partition(st)
         if (i == j)
            add_q(Q[i], succ) /* Insert succ at the tail of Q[i] */
            V = V[i] + {succ} /* Add succ to local State graph V[i] */
         else
            send(j, succ) /* Send the state st to its owner j */
  if (Q[i] != {})
    bfs_po(Q[i])
}
\end{verbatim}
Notice that a cycle containing state $s$ would be detected if there exist some successor of $s$ in the already generated local state graph $V[i]$ but such a successor is yet to be  expanded  OR the owner of representative of the successor i.e. $partition(t(s))$ is less that the id of the current node $i$.    
  
\subsection{A Distributed BFS Algorithm with both Partial Order and Symmetry Reductions}

We can extend the algorithm presented in the previous
section to handle both partial order and symmetry based reductions
in a distributed setting as follows:  

\begin{verbatim}
// Distributed BFS with Symmetry and POR

Search(i, Start_state) { ... }

visit(i) { ... }

check_C2*(s, X){
  for all (t in X)
     if ( [(rep(t(s)) is in V[i]) AND (t(s) is not in Q[i])] OR
          [partition(rep(t(s))) < i] )
        return False   
  return True
}

bfs_po_symm(Q, i) {
  s = pop_q(Q[i])

  for each (sequential process P)
    nxt := all transitions of P in ample(s)
    for each (t in nxt)
      succ = t(s)
      j = partition(succ)
      if(rep(succ) is not in V)
         if(j == i)
           add_q(Q[i], succ) /* Insert succ at the tail of Q */
           V[i] = V[i] + {rep(succ)} 
           /* Add rep(succ) to local State graph partition V[i] */
         else
           send(j, succ)
  if(Q[i] != {})
    bfs_po_symm(Q, i);
}
\end{verbatim}
Notice that a cycle would be detected if there exist some (representative) successor of $s$ in the already generated local state graph $V[i]$ but such a successor is yet to be  expanded  OR the owner of representative of the successor i.e. $partition(rep(t(s)))$ is less that the id of the current node $i$.

\end{document}